\documentclass[aps,pre,twocolumn,superscriptaddress,showpacs]{revtex4}
\usepackage[dvips]{graphicx}

\usepackage{amsmath}
\usepackage{amssymb}
\usepackage{tikz}
\usepackage{graphicx}
\usepackage{dcolumn}
\usepackage{bm}
\usepackage{epsfig}
\usepackage{psfrag}
\usepackage{xcolor}
\usepackage{color}
\def\II{\hbox{$1\hskip -1.2pt\vrule depth 0pt height 1.6ex width 0.7pt\vrule depth 0pt height 0.3pt width 0.12em$}}
\newcommand{\reffig}[1]{\mbox{Fig.~\ref{#1}}}
\newcommand{\refeq}[1]{\mbox{Eq.~(\ref{#1})}}
\newcommand{\refsec}[1]{\mbox{Sec.~\ref{#1}}}

\newcommand{\be}{\begin{equation}}
\newcommand{\ee}{\end{equation}}
\newcommand{\bal}{\begin{align}}
\newcommand{\eal}{\end{align}}
\newcommand{\ba}{\begin{eqnarray}}
\newcommand{\ea}{\end{eqnarray}}

\newcommand{\T}{${\mathcal T}\,$}
\newcommand{\Ti}{${\mathcal T}$}
\def\II{\hbox{$1\hskip -1.2pt\vrule depth 0pt height 1.6ex width 0.7pt\vrule depth 0pt height 0.3pt width 0.12em$}}

\begin{document}

\title{Distributions of the Wigner reaction matrix for microwave networks with  symplectic symmetry in the presence of absorption}
\author{Micha{\l} {\L}awniczak}
\email{lawni@ifpan.edu.pl}
\address{Institute of Physics, Polish Academy of Sciences, Aleja \ Lotnik\'{o}w 32/46, 02-668 Warsaw, Poland}
\author{ Afshin Akhshani}
\address{Institute of Physics, Polish Academy of Sciences, Aleja \ Lotnik\'{o}w 32/46, 02-668 Warsaw, Poland}
\author{ Omer Farooq}
\address{Institute of Physics, Polish Academy of Sciences, Aleja \ Lotnik\'{o}w 32/46, 02-668 Warsaw, Poland}
\author{Ma{\l}gorzata Bia{\l}ous}
\address{Institute of Physics, Polish Academy of Sciences, Aleja \ Lotnik\'{o}w 32/46, 02-668 Warsaw, Poland}
\author{Szymon Bauch}
\address{Institute of Physics, Polish Academy of Sciences, Aleja \ Lotnik\'{o}w 32/46, 02-668 Warsaw, Poland}
\author{Barbara Dietz}
\email{Dietz@lzu.edu.cn}
\address{Lanzhou Center for Theoretical Physics and the Gansu Provincial Key Laboratory of Theoretical Physics, Lanzhou University, Lanzhou, Gansu 730000, China}
\address{Center for Theoretical Physics of Complex Systems, Institute for Basic Science (IBS), Daejeon 34126, Korea}
\author{Leszek Sirko}
\email{sirko@ifpan.edu.pl}
\address{Institute of Physics, Polish Academy of Sciences, Aleja \ Lotnik\'{o}w 32/46, 02-668 Warsaw, Poland}

\date{\today}
\begin{abstract}
	We report on experimental studies of the distribution of the reflection coefficients, and the imaginary and real parts of Wigner's reaction ($K$) matrix employing open microwave networks with symplectic symmetry and varying size of absorption. The results are compared to analytical predictions derived for the single-channel scattering case within the framework of random matrix theory (RMT). Furthermore, we performed Monte Carlo simulations based on the Heidelberg approach for the scattering ($S$) and $K$ matrix of open quantum-chaotic systems and the two-point correlation function of the $S$-matrix elements. The analytical results and the Monte Carlo simulations depend on the size of absorption. To verify them, we performed experiments with microwave networks for various absorption strengths. We show that deviations from RMT predictions observed in the spectral properties of the corresponding closed quantum graph, and attributed to the presence of nonuniversal short periodic orbits, does not have any visible effects on the distributions of the reflection coefficients and the $K$ and $S$ matrices associated with the corresponding open quantum graph.
\end{abstract}

\maketitle

\section{Introduction\label{Intro}}
Quantum chaotic scattering was introduced  seventy years ago  to describe  properties of large complex quantum systems~\cite{Wigner1951,Haake2018,Weidenmueller2009}. Due to decoherence controllable experimental investigations of complex quantum systems are extremely difficult. A multitude of physical problems from the field of quantum chaos were experimentally tackled with the help of microwave networks and cavities simulating, respectively, quantum graphs~\cite{Hul2004,Lawniczak2010,Rehemanjiang2016}, and billiards~\cite{Stoeckmann1990,Sridhar1994,Sirko1997,Hlushchuk2000,Dietz2010,Zheng2006,Hul2005}. The present paper shows how microwave networks can be applied for the experimental study of properties of Wigner's reaction ($K$) matrix for quantum systems with symplectic symmetry. The experimental results are compared to exact analytical results for the single-channel scattering case~\cite{Fyodorov2004} and to Monte Carlo simulations based on random-matrix theory (RMT).

Quantum graphs consisting of one-dimensional wires connected by vertices were introduced more than 80 years ago by Linus Pauling~\cite{Pauling1936}. They are not only employed as basic mathematical objects~\cite{Kuchment2004,Gnutzmann2006,Berkolaiko2013} but are also indispensable in modeling of physical networks in the limit where the lengths of the wires are much bigger than their widths~\cite{Kottos1997,Hul2004}.  They have been used to simulate a large variety of systems and models, e.g., superconducting quantum circuits~\cite{Jooya2016}, quantum circuits in tunnel junctions~\cite{Namarvar2016} and realizations of high-dimensional multipartite quantum states~\cite{Krenn2017}. Quantum graphs consisting of bonds of incommensurable lengths provide invaluable tools for the study of closed~\cite{Gnutzmann2004} and open ~\cite{Kottos2000,Pluhar2014} quantum-chaotic systems. 

Because of the formal equivalence of the Schr\"odinger equation  describing quantum graphs and the telegraph equation of the corresponding microwave networks \cite{Hul2004,Lawniczak2010,Lawniczak2019b} the latter can be used to simulate the former. Indeed, microwave networks have been realized for the study of closed and open quantum-chaotic systems for all three Wigner-Dyson symmetry classes. Within random matrix theory (RMT) these are associated with the Gaussian ensemble (GE) of random matrices with corresponding universality class~\cite{Mehta1990}. If time-reversal (\Ti) invariance is conserved, the appropriate ensemble is the Gaussian orthogonal ensemble (GOE) for integer-spin systems~\cite{Hul2004,Lawniczak2008,Hul2012,Lawniczak2016,Dietz2017b,Lawniczak2019} and the Gaussian symplectic ensemble (GSE) for half-integer spin systems~\cite{Scharf1988,Dietz1990,Rehemanjiang2016,Lu2020,Che2021,Ramirez2022}. For systems  for which \T invariance is violated, it is the Gaussian unitary ensemble (GUE) \cite{Hul2004,Lawniczak2010,Lawniczak2019b,Bialous2016,Lawniczak2020,Che2022}.

Spectral properties of quantum graphs belonging to the symplectic universality class have been analyzed in microwave networks of corresponding geometry~\cite{Rehemanjiang2016,Rehemanjiang2018,Che2021}, also parametric ones~\cite{Lu2020}. In such experiments the scattering ($S$) matrix is measured as function of the microwave frequency. The $M\times M$ $K$ matrix, $\hat K$, of a scattering system with $M$ open channels is related to the associated $S$ matrix, $\hat S$, as~\cite{Fyodorov2005,Hemmady2006},
\begin{equation}
\label{Eq.1}
	\hat K =i(\hat S-\II_M)(\hat S+\II_M)^{-1},
\end{equation}
with $\II_M$ denoting the $M$-dimensional unit matrix. The importance of the $K$ matrix stems from the fact that it links the properties of the dynamics in the reaction region with the scattering processes observed in the asymptotic region, whereas the $S$ matrix provides a relation between the outgoing and incoming waves. But most importantly, the imaginary part of the $K$ matrix is proportional to that of the local Green function, which is known in solid-state physics as the local density of states (LDoS)~\cite{Mirlin2000,Fyodorov2004}. Furthermore, the electric impedance of a microwave  cavity, $\hat Z$, is directly related to the $K$ matrix, $\hat Z=i\hat K$~\cite{Hemmady2005,Zheng2006}. 

The $K$ matrix was studied experimentally in single-port measurements using microwave cavities~\cite{Mendez-Sanchez2003,Hemmady2005,Hemmady2006} and in microwave networks~\cite{Lawniczak2008,Hul2005a,Hul2007,Lawniczak2009} for GOE systems. Furthermore, distributions of the off-diagonal elements of the $S$ matrix of typical quantum-chaotic scattering systems were studied experimentally in~\cite{Dietz2010,Kumar2013,Kumar2017} in microwave cavities with partially violated \T invariance. The $S$ and $K$ matrix were investigated for complete violation of \T invariance with microwave networks in Ref.~\cite{Lawniczak2020}, in Ref.~\cite{Lawniczak2019b} for the case of large absorption. The situation is different for open chaotic systems with symplectic symmetry for which the $K$ matrix has not been studied so far. In this paper we report on elaborate experiments with microwave networks which were performed to close this gap.

In~\refsec{Sympl} we review general properties of closed and open quantum systems belonging to the symplectic universality class and structure of the associated Hermitian and unitary scattering matrices. We will specify the design of the quantum graphs with symplectic symmetry that are considered in the present paper. Then we will introduce the experimental setup in~\refsec{Exp}, the features of random matrices from the GSE and the corresponding random scattering matrix in~\refsec{RMT}. In~\refsec{ExpRes} we will present the results of the experiments and finally discuss them in~\refsec{Concl}.

\section{Salient features of \T invariant half-integer spin systems\label{Sympl}}

In this section we briefly review the properties of Hamiltonian systems which belong to the symplectic universality class. A detailed and comprehensive description can be found in~\cite{Haake2018}. The Hamiltonian is classified by its properties under application of the time-reversal operator $\hat T=\hat U\mathcal{C}$, where $\hat U$ is a unitary matrix and $\mathcal{C}$ denotes complex conjugation. For spinless particles the conventional time-reversal operation is complex conjugation, $\hat T=\mathcal{C}$ and $\hat T^2=1$. If \T invariance is violated, the associated Hamiltonian is complex Hermitian, $\hat H=\hat H^\dagger$, and thus belongs to the unitary symmetry class whereas, if it is preserved, $\hat T\hat H\hat T^{-1}=\hat H$, it becomes real symmetric in a \Ti-invariant basis and belongs to the orthogonal universality class. 

For half-integer particle systems with \T invariance $\hat T$ squares to -1, $\hat T^2=-1$, implying that, if $\psi$ is an eigenfunction of the Hamiltonian $\hat H$, then $\hat T\psi$ is also one for the same eigenvalue and is orthogonal to $\psi$, $\langle \psi\vert\hat T\psi\rangle =0$. Thus, the eigenvalues of $\hat H$ exhibit Kramer's degeneracy. Accordingly, we may choose a basis of the form $\mathcal{B}=\{\vert 1\rangle,\vert 2\rangle\dots,\vert N\rangle,\vert\hat T1\rangle,\vert \hat T2\rangle,\dots,\vert\hat T N\rangle\}$ for a $2N$-dimensional Hamiltonian. In this basis the $\hat T$ operator adopts the form
\be
\hat T=\hat Y\mathcal{C},\, \hat Y=\begin{pmatrix}\hat 0_N &-\II_N\\
\II_N&\hat 0_N\end{pmatrix},\label{Top}
\ee
where $\hat 0_N$ is the $N\times N$ dimensional zero matrix and the Hamiltonian can be written as
\be
\hat H=\begin{pmatrix}
\hat H_0 &\hat V\\ \hat V^\dagger &\hat H_1\end{pmatrix},\, \hat H_0=\hat H_0^\dagger,\, \hat H_1=\hat H_1^\dagger.\label{Ham0}
\ee
Here, $\hat H_0,\hat H_1,\hat V$ are $N\times N$ dimensional matrices. Time-reversal invariance implies that the Hamiltonian equals its symplectic transpose, $\hat H=\hat Y\hat H^T\hat Y^T$, with $\hat H^T$ denoting the transpose of $\hat H$, yielding that $\hat H_1$ is the complex conjugate of $\hat H_0$, $\hat H_1=\hat H_0^\ast$, and $\hat V^T=-\hat V$, i.e.,
\be
\hat H=\begin{pmatrix}
\hat H_0 &\hat V\\ -\hat V^\ast &\hat H_0^\ast\end{pmatrix},  \hat H_0=\hat H_0^\dagger,\, \hat V=-\hat V^T.\label{Ham}
\ee
Proceeding as described in Ref.~\cite{Haake2018} and rearranging the basis $\mathcal{B}$ defined above to $\mathcal{B}^q=\{\vert 1\rangle,\vert\hat T 1\rangle,\dots,\vert N\rangle,\vert\hat T N\rangle\}$, $\hat H$ can be written in the quaternion representation, that is, in terms of an $N\times N$ matrix whose matrix elements are $2\times 2$ quaternion matrices,
\be
\hat h_{mn}=h^{(0)}_{mn}\II_2 +\boldsymbol{h}_{mn}\cdot\boldsymbol{\tau},\, n,m=1,\dots ,N.
\label{quaternion}
\ee
Here, $\boldsymbol{\tau}=-i\boldsymbol{\sigma}$ with the components of $\boldsymbol{\sigma}$, $\sigma_i,\, i= 1,2,3$, referring to the three Pauli matrices. Time-reversal invariance implies that the matrices $\hat h_{nm}$ are quaternion real, $h^{(\mu)}_{mn}=h^{(\mu)\ast}_{mn}, \mu=0,\dots ,3$, and Hermiticity yields $h^{(0)}_{mn}=h^{(0)}_{nm}$, $\boldsymbol{h}_{mn}=-\boldsymbol{h}_{nm}$, and thus $\hat h_{nn}=h^{(0)}_{nn}\II_2$.

To generate random matrices for the GSE $\hat H_0$ is replaced by a random matrix from the GUE and, similarly, the entries of $\hat V$ are Gaussian distributed with zero mean and the same variance as the off-diagonal elements of $\hat H_0$. Equivalently, the matrix elements of $\hat h_{nm}$ in~\refeq{quaternion} are replaced by Gaussian distributed random numbers with zero mean, where the variance of the matrix elements $h_{nn}^{(0)}$ is by a factor of $\sqrt{2}$ larger than that of $h_{nm}^{(\mu)},n\ne m$. 

The design of the quantum graphs used in the present paper was chosen such that the corresponding Hamiltonian~\cite{Rehemanjiang2016,Lu2020} attains the form given in~\refeq{Ham}. Accordingly, they are constructed from two quantum graphs with the same geometry belonging to the unitary symmetry class, referred to as GUE graphs in the following. Each GUE graph consists of $\mathcal{V}$ vertices, where corresponding vertices are denoted by $i$ and $i^\prime$ with $i,i^\prime=1,\dots,\mathcal{V}$. In the notation introduced above $i$ and $i^\prime$ correspond to $\vert i\rangle$ and $\vert\hat Ti\rangle$, respectively.  The vertices are connected by bonds according to the connectivity matrix $\hat C$, which has vanishing diagonal elements $C_{\tilde i\tilde i}=0$ and off-diagonal elements $C_{\tilde i\tilde j}=1$ if vertices $\tilde i$ and $\tilde j$ are connected and $C_{\tilde i\tilde j}=0$ otherwise. Here, $\tilde i=i$ for $1\leq\tilde i\leq\mathcal{V}$, and $\tilde i=\mathcal{V}+i$ corresponds to $i^\prime$ for $\mathcal{V}+1\leq\tilde i\leq 2\mathcal{V}$. Corresponding bond lengths coincide, $L_{ij}=L_{i^\prime j^\prime}$. In our realizations the GUE graphs are connected at four vertices, e.g., at $i_0,j_0^\prime$ and  $j_0,i_0^\prime$ through bonds with same lengths $L_{i_0 j_0^\prime}=L_{j_0 i_0^\prime}$. In order to comply with the requirement $\hat V=-\hat V^T$, that is, $V_{i_0 j_0^\prime}=-V_{j_0 i_0^\prime}$, an additional phase of $\pi$ is introduced on one of the bonds, e.g., on that connecting $i_0$ and $j_0^\prime$. The wave function components $\psi_{\tilde i\tilde j}(x)$ on the bonds are solutions of the one-dimensional Schr\"odinger equation
\be
\left(-i\frac{d}{dx}-A_{\tilde i\tilde j}\right)^2\psi_{\tilde i\tilde j}(x)+k^2\psi_{\tilde i\tilde j}(x)=0,\label{WEB}
\ee
where $\hat A=-\hat A^T$ denotes the magnetic vector potential which induces \Ti-invariance violation, and $A_{ij}=-A_{i^\prime j^\prime}$ on corresponding bonds in $\hat H_0$ and $\hat H^\star_0$, respectively. The coordinate $x$ varies along the bond from $x=0$ at vertex $\tilde i$ to $x=L_{\tilde i\tilde j}$ at vertex $\tilde j$. On the bonds that couple the GUE graphs the Schr\"odinger equation~(\ref{WEB}) with $A_{\tilde i\tilde j}=0$ applies. 

The wave-function components are subject to boundary conditions imposed at the vertices that ensure continuity and conservation of the current~\cite{Kottos1999}. We restrict here to Neumann boundary conditions, which can be modeled experimentally with microwave networks~\cite{Hul2004}. They constitute a special case of $\delta$-type boundary conditions~\cite{Kottos1999,Kostrykin1999,Texier2001,Kuchment2004}. The eigenwavenumbers of quantum graphs with these boundary conditions are determined by solving the equation~\cite{Kottos1999},
\begin{equation}
\det\hat h(k)=0,
\end{equation}
with
\begin{equation}
h_{\tilde i\tilde j}(k)=\left\{{\begin{array}{cc}
        -\sum_{\tilde m\ne\tilde i}\cos\left(kL_{\tilde i\tilde m}\right)\frac{C_{\tilde i\tilde m}}{\sin\left(kL_{\tilde i\tilde m}\right)}&, \tilde i=\tilde j\\
	C_{\tilde i\tilde j}e^{-iA_{\tilde i\tilde j}L_{\tilde i\tilde j}-i\varphi_{\tilde i\tilde j}}
	\left[\sin(kL_{\tilde i\tilde j})\right]^{-1}&, \tilde i\ne\tilde j\\
        \end{array}}
        \right. ,\label{QuantE}
\end{equation}
where $\varphi_{\tilde i\tilde j}=\pi$ for $(\tilde i=i_0, \tilde j=j_0^\prime$) and zero otherwise. For the magnetic vector potential we chose $\vert A_{\tilde i\tilde j}\vert =\frac{\pi}{2}$ on some or all of the bonds of the GUE graphs. The choice of the sign of $A_{\tilde i\tilde j}$, namely $A_{ij}=-A_{i^\prime j^\prime}$ on corresponding bonds of the two GUE graphs ensures that the associated submatrices are complex conjugate to each other. The components of the associated eigenvectors yield the values of the wave functions at the vertices and thus the eigenfunctions~\cite{Kottos1999}. 

A schematic view of a realization of a GSE graph is shown in the inset of~\reffig{Fig1}. There, $i_0=1$, $i_0^\prime=1^\prime$ and $j_0=2$, $j_0^\prime =2^\prime$ and the vector potentials $\hat A$ introduced on the two GUE graphs are indicated by the phases $+\frac{\pi}{2}$, respectively $-\frac{\pi}{2}$. In Refs.~\cite{Gnutzmann2004,Pluhar2014} it was proven rigorously, that quantum graphs with Neumann boundary conditions generally exhibit spectral properties of a typical wave-chaotic system if the bond lengths are incommensurate. 

In Refs.~\cite{Rehemanjiang2016,Lu2020,Che2021} experiments with microwave networks that model such quantum graphs yielded that their spectral properties are close to those of random matrices from the GSE. Deviations from the RMT predictions, also observed in numerical studies where much longer eigenvalue sequences are available, can have two origins. First, in the quantum graphs under consideration the GUE graphs are coupled through two bonds only. To convince oneself that random matrices of the form~\refeq{Ham} with rank of $\hat V$ less than $N$ exhibit a level repulsion $\propto s^4$ one can proceed as in~\cite{Berry1988,Haake2018,Rehemanjiang2018} to derive the corresponding Wigner surmise, that is, the nearest-neighbor spacing distribution for $N=2$, which was shown for the GSE to agree well with that of random matrices from the GSE~\cite{Dietz1990}. Actually, it has been demonstrated in~\cite{Dietz2006a} that the spectral properties and properties of the eigenvectors of random matrices from the GOE or GUE of the form~\refeq{Ham0} that are coupled by a rank 1 and a full perturbation matrix $\hat V$, respectively, agree well. Second, quantum graphs with Neumann boundary conditions comprise eigenfunctions that are localized on a fraction of them. Their contributions could be extracted explicitly experimentally and numerically for parametric GSE graphs in~\cite{Lu2020}. 

Furthermore, it has been demonstrated in Refs.~\cite{Pluhar2013,Pluhar2013a,Pluhar2014} that the correlation functions of the $S$-matrix elements of open quantum graphs agree with RMT predictions~\cite{Verbaarschot1985,Fyodorov2005} for quantum chaotic scattering systems. The motivation of the present paper was to test the applicability of RMT for quantum chaotic scattering systems to open GSE graphs and to investigate the effect of such nonuniversal features on fluctuation properties of the $S$ matrix. Open quantum graphs are realized by attaching leads, that couple the graph to the environment, to corresponding vertices of the two GUE graphs. The $S$ matrix of a quantum graph with $\tilde  M=2M$ open channels can be brought to the form~\cite{Kottos1999}
\begin{align}
\label{SVertex}
	\hat S_{\mathcal{V}}(k)&=2i\hat W^T\left[\hat h(k)+\frac{i}{2}\hat W\hat W^T\right]^{-1}\hat W-\II_{\tilde M}\\
	\nonumber &=\left[i\hat W^T\hat h^{-1}(k)\hat W-\II_{\tilde M}\right]\left[i\hat W^T\hat h^{-1}(k)\hat W+\II_{\tilde M}\right]^{-1},
\end{align}
which is similar in form to that derived on the basis of the $S$-matrix formalism for compound nucleus reactions~\cite{Mahaux1969}, see ~\refsec{RMT}. Here, the $\tilde M\times 2\mathcal{V}$-dimensional matrix $\hat W$ is the coupling matrix which accounts for the coupling of leads to $\tilde M$ vertices and $\hat h(k)$ is given in the basis $\mathcal{B}$. We chose $\tilde M=2$ and attached leads to the ports marked by $P1$ and $P1^\prime$ in~\reffig{Fig1} at the vertices marked by $3,3^\prime$, that is, $\hat W_{P\tilde i\tilde j}=1$ for $\tilde i=1, \tilde j=3$, and $\tilde i=1^\prime,\tilde j=\mathcal{V}+3$ and zero otherwise. 

The $S$ matrix is \T invariant if $\hat T\hat S_\mathcal{V}\hat T^{-1}=\hat S_\mathcal{V}^\dagger$ with $\hat T$ defined in~\refeq{Top}, that is, when the $S$ matrix equals its inverse, $\hat S_\mathcal{V} =\hat S_\mathcal{V}^{-1}$. Applying the $\hat T$-operator to $\hat S_{\mathcal V}$ in the presentation given in the second line of~\refeq{SVertex} shows, that this indeed is the case for the design under consideration since the $\hat T$ operator commutes with $\hat h(k)$ and also with $\hat W$, which is a real matrix that complies with the form~\refeq{Ham}. Like in the Hermitian case the unitary $S$ matrix can be written in the quaternion representation with basis $\mathcal{B}^q$~\cite{Haake2018},
\be
\hat s_{mn}=s^{(0)}_{mn}\II_2 +\boldsymbol{s}_{mn}\cdot\boldsymbol{\tau}.
\label{Squaternion}
\ee
Time-reversal invariance implies that $s^{(0)}_{mn}=s^{(0)}_{nm}$ and $\boldsymbol{s}_{mn}=-\boldsymbol{s}_{nm}$ where, in distinction to the Hermitian case, the coefficients   $s^{(0)}_{mn},\boldsymbol{s}_{mn}$ are in general complex numbers. For the case $\tilde M=2$, i.e., $M=1$, the antisymmetry property implies that the S matrix is diagonal,
\be
\hat S=\begin{pmatrix}S_{P1P1}&S_{P1P1^\prime}\\S_{P1^\prime P1}&S_{P1^\prime P1^\prime}\end{pmatrix}=s^{(0)}_{11}\II_2,\label{Smatrix}
\ee
that is, $S_{P1P1}=S_{P1^\prime P1^\prime}=s_{11}$, yielding for the $K$ matrix~\refeq{Eq.1}
\be
K_{P\tilde iP\tilde j}=i\frac{\left[s_{11}-1\right]}{\left[s_{11}+1\right]}\delta_{P\tilde iP\tilde j}.\label{Kmatrix}
\ee
Thus, the $S$-matrix elements can be expressed as $\hat S_{P\tilde iP\tilde i}\equiv s_{11}=re^{i\theta}=\sqrt{R}e^{i\theta}$, where $r$, $R$ and $\theta$ are the scattering amplitude, reflection coefficient and the phase measured at the port. 

\section{Experimental setup\label{Exp}} 
\begin{figure}[h!]
\includegraphics[width=\linewidth]{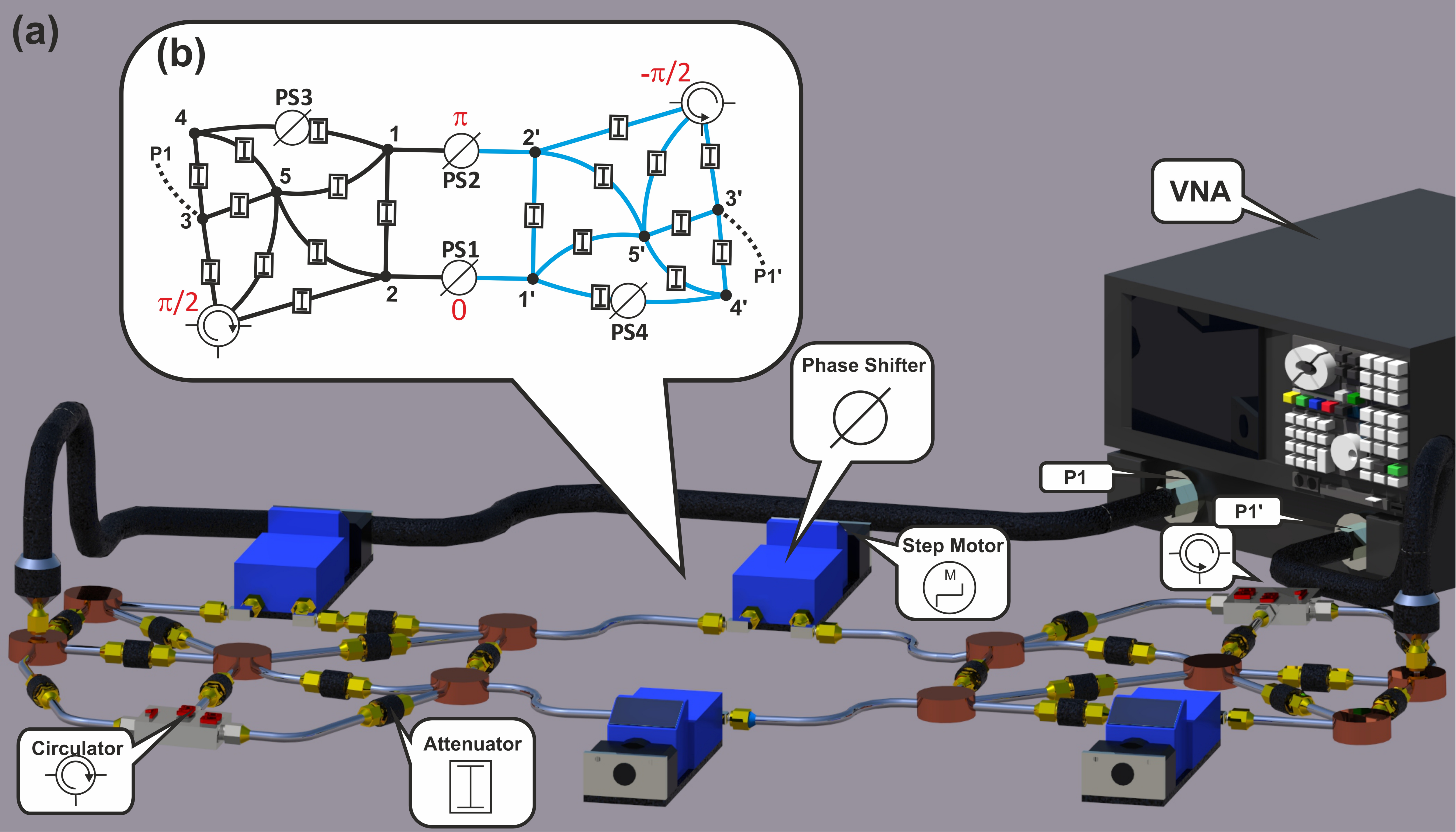}
	\caption{Three-dimensional (a) and two-dimensional (b) schemes of the microwave network with symplectic symmetry. It is constructed from two geometrically identical GUE subgraphs, marked black and turquoise in (b). Time-reversal invariance violation is induced by T-shaped circulators of opposite orientation introduced at corresponding vertices, that lead to an additional phase $\frac{\pi}{2}$, respectively $-\frac{\pi}{2}$, as indicated in~(b). The subgraphs are connected at two vertex pairs marked by $(1,2^\prime)$ iand $(2,1^\prime)$ through coaxial cables that comprise phase shifters (PS1 and PS2) that induce a relative phase $\pi $ of the waves traveling through them. Different realizations of the GSE graph were obtained by increasing the lengths of two corresponding bonds with phase shifters (PS3 and PS4) by the same amount. The absorption strength was changed by introducing 20 1~dB or 2~dB attenuators.}\label{Fig1}
\end{figure}

In the experiments quantum graphs with symplectic symmetry were modeled with microwave networks~\cite{Rehemanjiang2016,Lu2020,Che2021} with symplectic symmetry comprising two GUE microwave networks with identical geometry~\cite{Lawniczak2010}. Their bonds are plotted in black and turquoise in~\reffig{Fig1} (b). Corresponding vertices are numbered by $j$ and $j^\prime$. Two AEROTEK microwave circulators~\cite{Lawniczak2010}  with low insertion loss are introduced with opposite orientation at corresponding vertices. They operate in the frequency range $\nu \in [3.5-7.5]$~GHz and cause phase shifts $\pm\frac{\pi}{2}$ and thus induce \Ti-invariance violation~\cite{Kottos1999}~\footnote{The vertex scattering matrix associated with a circulator depends on frequency~\cite{Berkolaiko2013,Exner2018,Lawniczak2020,Che2022} and thus the microwave network at most models a GUE quantum graph~\cite{Kottos1999}. However, it has been shown in numerous experiments that it exhibits the same features and, most importantly, wave-dynamical chaos that complies with GUE statistics.}. Circulators are non-reciprocal three-port passive devices, that allow waves entering at port $1,2$ or $3$, to exit through, respectively, $2,3$ or $1$. The two parts of the microwave network are connected by two coaxial cables of same length, but microwaves traveling through them have a relative phase $\pi$ which is realized by the phase shifters marked by PS1 and PS2 in~\reffig{Fig1} (b), to enforce the appearance of Kramer's doublets. Then, transmission  from port $P1$ to port $P1^\prime$ or vice versa through the connecting bonds is suppressed due to destructive interference, where the ports correspond to leads in the associated quantum graph. Accordingly, these experiments effectively are single-port measurements simulating single-channel scattering, as expected for the two-dimensional $S$ matrix of a GSE graph according to~\refeq{Smatrix} leading to the $K$ matrix~\refeq{Kmatrix}. Note, that in the experiments $S_{P1P1}$ and $S_{P1^\prime P1^\prime}$ slightly differ from each other. The two-port $S$ matrix of the microwave network was measured using a vector network analyzer (VNA), Agilent E8364B. The network was connected to the VNA through HP 85133-616 and HP 85133-617 flexible microwave cables. Ensembles of networks were realized by changing by the same amount the lengths of two corresponding bonds of the network using the phase shifters marked by PS3 and PS4 in~\reffig{Fig1}. To attain absorption 1~dB and 2~dB attenuators were introduced into the microwave network. Note that, in particular, for reflection measurements, which yield the diagonal elements of the $S$ matrix, direct processes need to be eliminated to obtain the $S$-matrix elements for perfect coupling to the VNA. This is carried out in microwave networks by employing the impedance approach~\cite{Hemmady2005,Lawniczak2010}.

\begin{figure}[h!]
\includegraphics[width=1.0\linewidth]{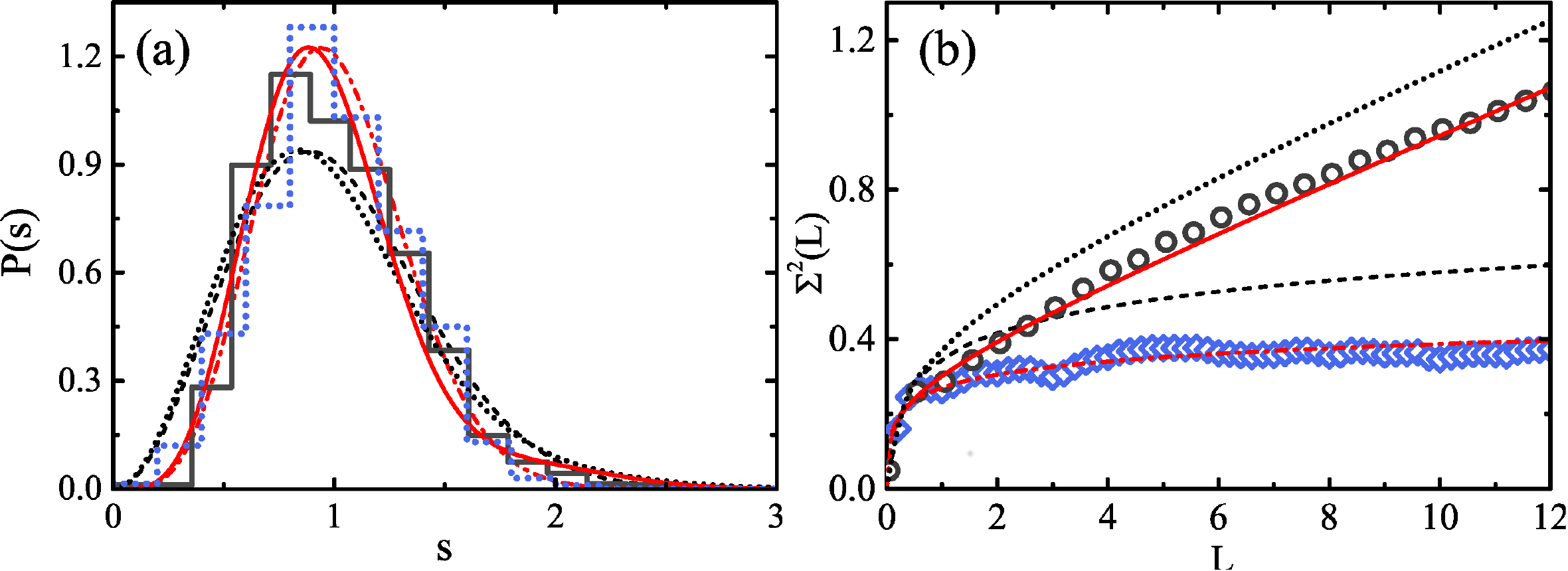}
	\caption{(a) Nearest-neighbor spacing distribution $P(s)$. The histogram exhibits the experimental distribution. A fraction $\varphi=0.94$ of the eigenfrequencies could be identified. The results are compared with the distributions obtained from missing-level statistics for the GSE (red full line) and GUE (black dotted line), and to those for the GSE (red dotted-dashed lines) and GUE (black dashed lines). (b) The number variance $\Sigma^2 (L)$. The experimental results are plotted as black circles. Otherwise same as in (a). We also show the results for the complete sequence of 1900 eigenvalues of the corresponding quantum graph, which was computed numerically (blue-dotted histogram and diamonds).  
}\label{Fig2}
\end{figure}

In order to confirm that the network exhibits the properties of a graph with symplectic symmetry, we analyzed fluctuation properties in the eigenfrequency spectra of the microwave networks without attenuators. For this we rescaled the ordered eigenfrequencies $\nu_j$ to mean spacing unity by replacing them by the smooth part of the integrated spectral density given by Weyl's formula, $e_{j}=2\mathcal{L} \nu_j /c$~\cite{Weyl1913,Kottos1999,Rehemanjiang2016}, where the total optical length $\mathcal{L}$ of the network was varied from 7.09-7.28~m and $c$ is the speed of light. Note, that in the experiments only one of the Kramer doublet partners was identified and, accordingly, half of the total length was used for the rescaling. In~\reffig{Fig2} (a) we show the experimental results for the nearest-neighbor spacing distribution $P(s)$ (black histogram) and in (b) the number variance $\Sigma^2(L)$ (red circles)~\cite{Mehta1990}. Here, $\Sigma^2(L)=\left\langle\left[n(L)-\langle n(L)\rangle\right]^2\right\rangle$ with $n(L)$ denoting the integrated spectral density, that is the number of unfolded eigenfrequencies $e_j$ in an interval of length $L$. The averaging $\langle \cdot\rangle$ comprises a spectral average over 133 eigenfrequencies in each spectrum, determined in a frequency range $[3.5,7.5]$~GHz, where absorption, which hampers the determination of the eigenfrequencies~\cite{Che2021}, is sufficiently small~\cite{Bialous2017}, and an ensemble average over 21 quantum-graph realizations. Note, that for proper unfolding $\langle n(L)\rangle=L$. The curves clearly differ from those for the GSE (red dashed-dotted lines) and the GUE (black dashed lines). The reason is, that according to Weyl's formula $6\%$ of the eigenfrequencies are missing, yielding a fraction $\varphi=0.94$ of identified ones. Therefore, we compare the experimental curves with those obtained from a RMT model for missing levels (red full line)~\cite{Bohigas2004,Che2021,Bialous2016}. To illustrate that the spectral properties unambiguously agree with those of the GSE with $6\%$ missing levels, we added those for the GUE ( black dotted lines) with the same percentage of missing levels. Since the determination of complete sequences of eigenfrequencies was impossible we computed 1900 eigenvalues for the corresponding GSE quantum graph. The good agreement between the experimental and theoretical results corroborates that the microwave networks exhibit GSE symmetry. The nearest-neighbor spacing distribution and number variance are shown as  blue-dotted histogram and blue diamonds, respectively. Small deviations from the RMT prediction observed in $P(s)$ and $\Sigma^2(L)$ for the experimental and numerical curves have been shown in Ref.~\cite{Dietz2017b,Lu2020} to originate from the contribution of nonuniversal orbits that are confined to a fraction of the quantum graph.  

\section{Random matrix theory results\label{RMT}} In this paper analytical results for the distributions of the imaginary part $v=-\textrm{Im} \, K$ and the real part $u=\textrm{Re} \,K$ of a single-channel $K$ matrix~\refeq{Kmatrix} are tested for intermediate and large absorption, achieved by introducing 1~dB and 2~dB attenuators into the microwave network (see~\reffig{Fig1}). In Ref.~\cite{Fyodorov2004} an analytic expression is derived for quantum-chaotic scattering systems with symplectic symmetry for the distribution $P_0(x)$ of $x=\frac{1+R}{1-R}=\frac{u^2+v^2+1}{2v}$ for the single-channel scattering case~\refeq{Smatrix} in presence of absorption,
\begin{align}
\label{Eq.2}
	P_0(x) &=\frac{1}{2}\Bigl[A\gamma (x+1)+B\Bigr]e^{-\gamma(x+1)}\\
	\nonumber &+C(x,\gamma)e^{-\gamma x}\int_0^{\gamma}dt\frac{{\rm sinh}t}{t}.
\end{align}
Here, $A=e^{2\gamma}-1$, $B=1+2\gamma-e^{2\gamma}$, $C(x,\gamma)=\frac{\gamma^2(x+1)^2}{2}-\gamma(\gamma+1)(x+1)+\gamma$. The corresponding analytical results for the reflection coefficient $R$, the amplitude $r=\sqrt{R}$ and the real and imaginary parts of $K=u-iv$ are derived from that one, yielding
\begin{equation}
\label{Eq.3}
P(R)=\frac{2}{(1-R)^2}P_0\Bigl(\frac{1+R}{1-R}\Bigr),
\end{equation}
\begin{equation}
\label{Eq.4} P(v)=\frac{\sqrt{2}}{\pi
v^{3/2}}\int^{\infty}_{0}dqP_0\Bigl[q^2+\frac{1}{2}\Bigl(v+\frac{1}{v}\Bigr)\Bigr],
\end{equation}
\begin{equation}
\label{Eq.5} P(u)=\frac{1}{2\pi
\sqrt{u^{2}+1}}\int^{\infty}_{0}dqP_0\Bigl[\frac{\sqrt{u^2+1}}{2}\Bigl(q+\frac{1}{q}\Bigr)\Bigr],
\end{equation}
and $P(r)=2rP(R)$. For each realization of the microwave networks the absorption parameter $\gamma$ was determined by fitting the theoretical mean reflection coefficient $\langle R\rangle = \int _0^1dRRP(R)$ to the experimental ones, and similarly the analytical curve for $P(r)$ to the experimental distributions. Note, that in the experiments $S_{P1P1}$ and $S_{P1^\prime P1^\prime}$ slightly differ from each other. Therefore, we used their averages, $\langle R\rangle=\frac{1}{2}\left\{\langle\vert S_{P1P1}\vert^2\rangle +\langle\vert S_{P1^\prime P1^\prime}\vert^2\rangle\right\}$. Here, $\langle\cdot\rangle$ means spectral averaging over different frequency intervals and ensemble averaging over all graph realizations. The thereby determined values are $\gamma = 5.7 \pm 0.1$ for the network with 1~dB attenuators and $\gamma  =12.8\pm 0.2$ for the one with 2~dB attenuators. We also compare the experimental results to the analytical results for the GUE, which are obtained from $P_0(x)$ defined in~\refeq{Eq.2} by setting $C(x,\gamma)=0$ and replacing $\gamma$ by $\frac{\gamma}{2}$,
\be
\label{Eq.2s}
P_0(x) =\frac{1}{2}\Bigl[(e^{\gamma}-1)\gamma\frac{(x+1)}{2}+(1+\gamma-e^{\gamma})\Bigr]e^{-\frac{\gamma}{2}(x+1)}.
\ee
	Comparison with~\refeq{Eq.2} shows that due to the differing decay behavior of $P_0(x)$ for the GUE and GSE cases the corresponding  distributions obtained from Eqs.~(\ref{Eq.3})-(\ref{Eq.5}) should be well distinguishable for moderate values of $\gamma$. Furthermore, the average values $\langle r\rangle$ and $\langle R\rangle$ differ considerably for the GUE and GSE for a given value of $\gamma$. Yet, after rescaling of the reflection coefficients $R$ and amplitudes $r$ to average value unity, $\tilde R=\frac{R}{\langle R\rangle}$ and $\tilde r=\frac{r}{\langle r\rangle}$, their distributions will approach an exponential and bivariate Gaussian, respectively,
\be
P(\tilde R)\xrightarrow{\gamma\to\infty}e^{-\tilde R},\, P(\tilde r)\xrightarrow{\gamma\to\infty}\frac{\pi}{2}\tilde re^{-\frac{\pi}{4}\tilde r^2}
\label{Ericson}
\ee
in the Ericson regime~\cite{Ericson1960,Ericson2016}, so that they become indistinguishable. 
\begin{figure}[h!]
\includegraphics[width=\linewidth]{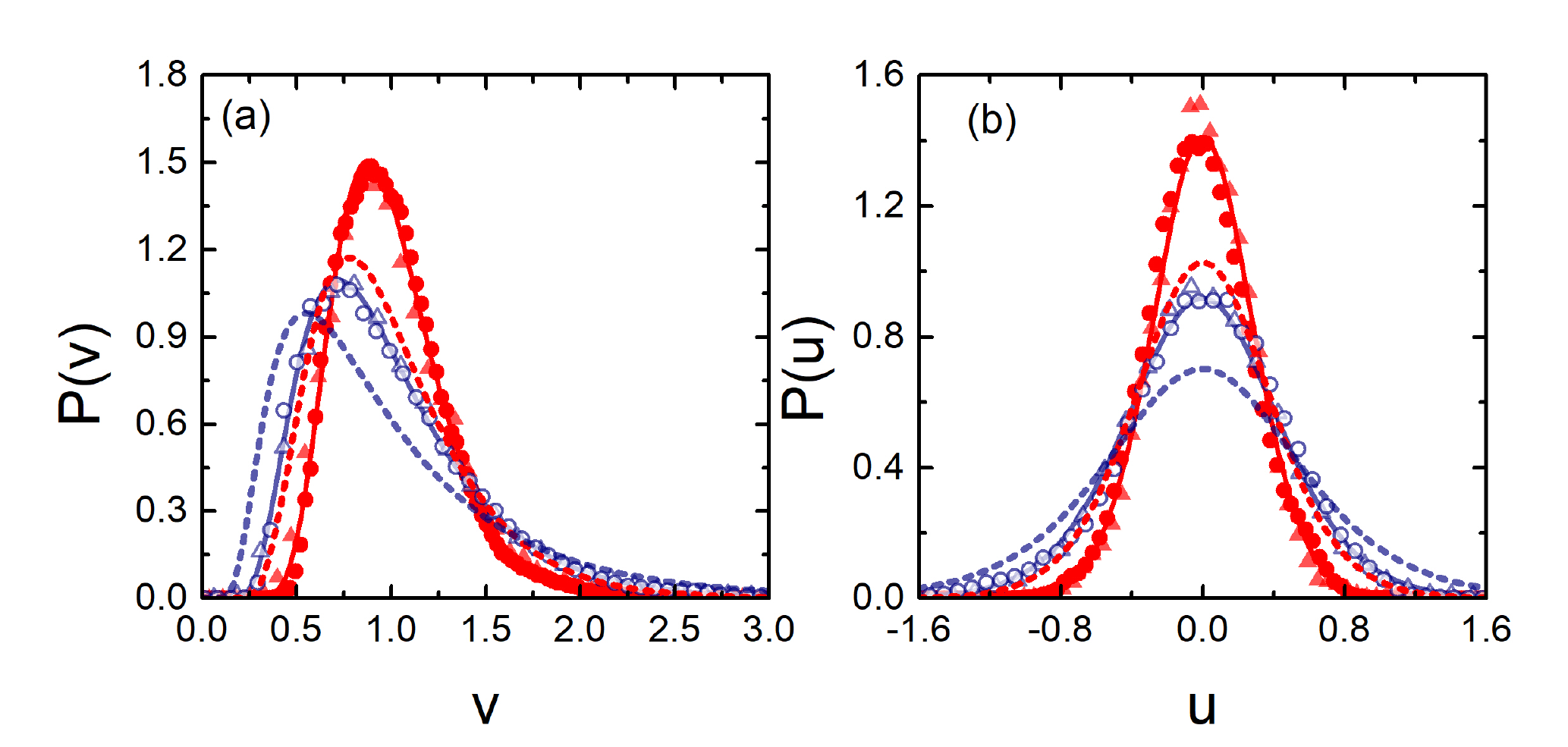}
	\caption{Distributions of (a) the imaginary part $P(v)$ and (b) the real part $P(u)$ of $K$ evaluated experimentally for the microwave network with symplectic symmetry and 1~dB attenuators (blue circles) respectively, 2~dB attenuators (red dots). The experimental results are compared with the analytical results Eqs.~(\ref{Eq.3})-(\ref{Eq.5}) for $\gamma  =5.7$ (blue solid line) and $\gamma  =12.8$ (red solid line). The corresponding RMT results are marked by blue open triangles and red full triangles for $\gamma  =5.7$ and $\gamma  =12.8$, respectively. The analytical results for the GUE are shown for the same values of $\gamma$ as dashed lines with corresponding color.
}\label{Fig3}
\end{figure}
\begin{figure}[h!]
\includegraphics[width=\linewidth]{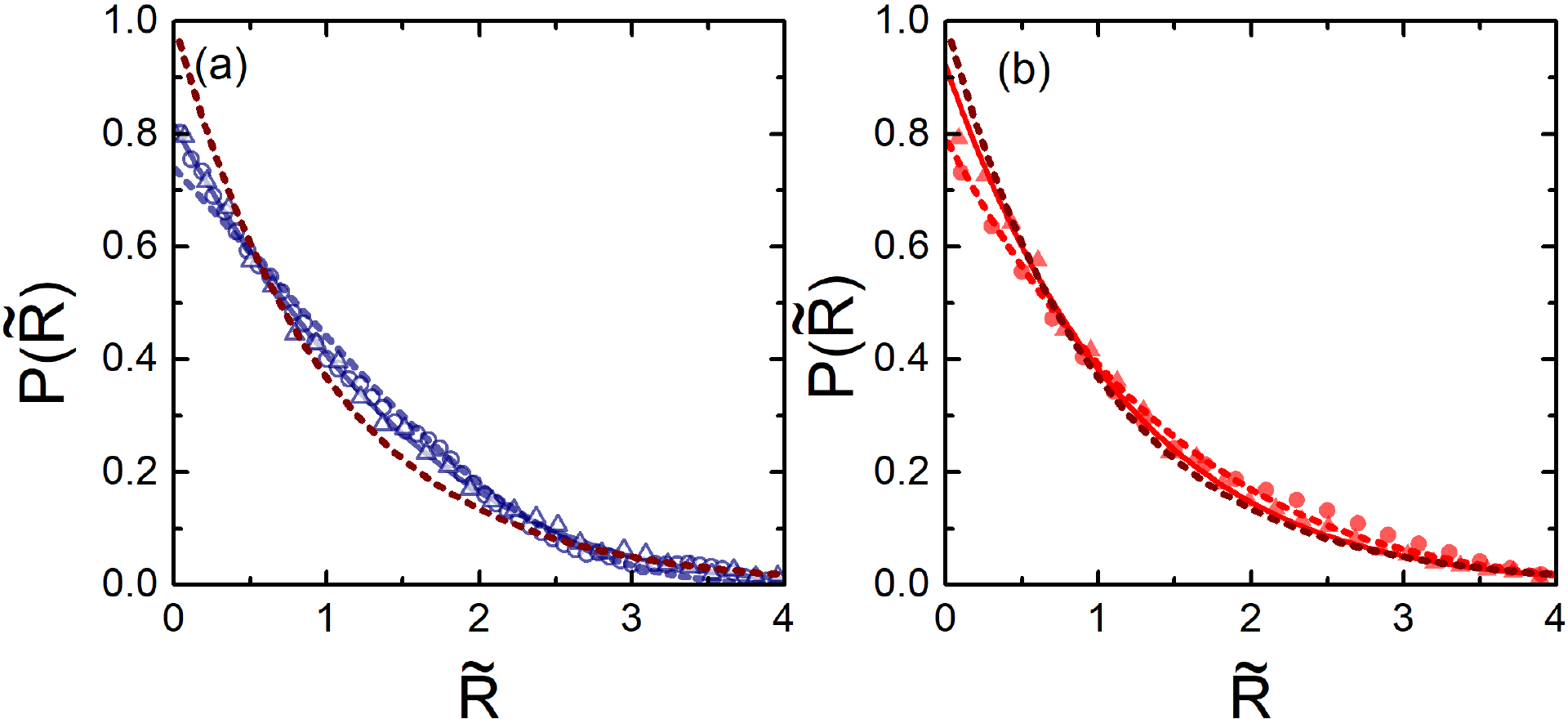}
	\caption{Same as~\reffig{Fig3} for the distributions of the rescaled reflection coefficients $P(\tilde R)$. Furthermore, we added the limitting curve for large absorption, i.e., for the Ericson regime as maroon dashed lines. 
}\label{Fig4}
\end{figure}

We performed Monte Carlo simulations using the $S$-matrix formalism~\cite{Mahaux1969} which was developed by Mahaux and Weidenm{\"u}ller in the context of compound nuclear reactions and employed for the derivation of exact analytical results for fluctuation properties of the $S$ matrix associated with a quantum-chaotic scattering system~\cite{Verbaarschot1985,Fyodorov2005,Dietz2009,Dietz2010,Kumar2013,Kumar2017},
\begin{equation}
	S(k)_{P\tilde jP\tilde l} = \delta_{P\tilde jP\tilde l} - i\sum_{\mu,\nu=1}^{2N}\hat W_{\mu P\tilde j}\left[\left(k\II-\hat H^{eff}\right)^{-1}\right]_{\mu\nu}\hat W_{\nu P\tilde l},
\label{MW}
\end{equation}
which is similar in form to the scattering matrix for open quantum graphs~\refeq{SVertex}.
In the microwave network $P\tilde j$ and $P\tilde l$ refer to antenna ports, and $\hat H^{eff}=\hat H-\frac{i}{2}\hat W\hat W^T$. Here, $\hat H$ denotes the $k$-independent random matrix from the appropriate GE and $k\II-\hat H$ mimicks the spectral fluctuation properties of the $k$-dependent Hamiltonian $\hat h(k)$ of the closed quantum graph or microwave network with no coupling to the environment. To model the properties of a graph belonging to the symplectic universality class it is replaced by a random matrix from the GSE. We use the quaternion representations Eqs.(~\ref{quaternion}) and~(\ref{Squaternion}) for $\hat H$ and $\hat S$. As outlined in~\refsec{Sympl} the entries of the quaternion matrices correspond in the quantum-graph Hamiltonian~\cite{Kottos1999,Rehemanjiang2016} to one vertex in one part of the GSE graph, one in the other one, and their coupling. The matrix elements of $\hat W$ are also given in the quaternion representation. They describe the coupling of the modes in the microwave networks to the environment via the antenna ports. Furthermore, absorption is modeled in the Monte Carlo simulations~\cite{Dietz2009,Dietz2010} by $2\Lambda$ fictitious channels. To ensure that the $2(\Lambda +1)\times 2N$-dimensional coupling matrix $\hat W$ complies with the symplectic properties of $\hat H$, we diagonalized a random matrix from the GSE and chose $\Lambda +1 < N$ of its eigenvectors, given in the quaternion basis, to generate $\hat W$. Then the orthogonality property holds for $\hat W$, that is, only the diagonal entries $w_n^2, n=1,\cdots,\Lambda +1$ of $\hat W^T\hat W$ are nonvanishing. This is in accordance with the property that the frequency-averaged $S$ matrix was diagonal in all microwave-network realizations after extracting direct processes as described above. The quantities $w_n$ are the input parameters of the RMT model~\refeq{MW} through the transmission coefficients
\be
T_{n} = 1 - \vert\left\langle{s_{nn}}\right\rangle\vert^2=\frac{4\pi^2w^2_{n}/d}{(1+\pi^2w^2_{n}/d)^2},
\label{Transm}
\ee
with $d$ denoting the mean resonance spacing. They provide a measure for the unitarity deficit of the average $S$ matrix~\cite{Dietz2010}. 

The transmission coefficients associated with an antenna port $P1$ in one subgraph and the corresponding one, $P1^\prime$, in the other one are determined from the measured reflection spectra after extraction of direct processes yielding with~\refeq{Transm} $T_{P1}\equiv T_{P1^\prime}\simeq 0.95$ for all measurements, as expected for quantum graphs with symplectic symmetry. Those related to the fictitious channels, accounted for through the parameter $\tau_{abs}=2\Lambda T_{f}$~\cite{Dietz2009}, are determined by fitting the RMT results for the normalized two-point correlation function of the $S$-matrix elements,
\begin{equation}
        C_{P\tilde iP\tilde i}(\epsilon) = 
	\frac{\langle S_{P\tilde iP\tilde i}(e)\, S_{P\tilde iP\tilde i}^\ast(e+\epsilon) \rangle - | \langle S_{P\tilde iP\tilde i}(e) \rangle |^2} 
	{\langle\vert S_{P\tilde iP\tilde i}(e)\vert^2 \rangle - | \langle S_{P\tilde iP\tilde i}(e) \rangle |^2}, 
	\label{AutoCorr}
\end{equation}
with $\tilde i=1,1^\prime$, to the measurement results at either of the two antenna ports and, similarly, the distribution of the amplitudes $\vert S_{P\tilde iP\tilde i}\vert$ to the experimental ones. We thereby confirmed the values obtained from the reflection coefficients as described above, thus corroborating that the complexity of the wave dynamics in the two parts of the microwave networks and their coupling through just two bonds suffices to generate agreement of the fluctuation properties of the $S$ matrix with those of a quantum-chaotic scattering system. 

\section{Experimental results\label{ExpRes}} The experimental effort to realize microwave networks that are suitable for the study of the properties of the $K$ and $S$ matrix and comparison with RMT prediction is facilitated when adding a small absorption to each cable, because it reduces the frequency dependence of the resonance parameters which is presupposed in the RMT approach. All experimental distributions are obtained by averaging over 41 and 81 realizations of the networks containing 1~dB and 2~dB attenuators, yielding total optical lengths from 7.09 m to 7.28 m and from 6.89 m to 6.92 m, respectively. Furthermore, we avaraged over the results obtained from the measurements at each of the two antenna ports. In~\reffig{Fig3} (a) and (b) are shown the experimental results for the distributions of the imaginary part, $P(v)$, and the real part, $P(u)$, of $K$, in~\reffig{Fig4} (a) and (b) those for the distributions of the rescaled reflection coefficients, $P(\tilde R)$, for the microwave networks with 1~dB (blue circles) and 2~dB (red dots) attenuators, respectively. The analytical results for $\gamma =5.7, 12.8$ are exhibited as solid lines, the corresponding RMT results as blue open and red full triangles. We also include the analytical results for the GUE for the same values of $\gamma$ as dashed lines. They clearly differ from the experimental and analytical results for the GSE. For $P(\tilde R)$ the curve for the case with 2~dB absorption is close to that for the Ericson regime~\refeq{Ericson}.

\begin{figure}[h!]
\includegraphics[width=\linewidth]{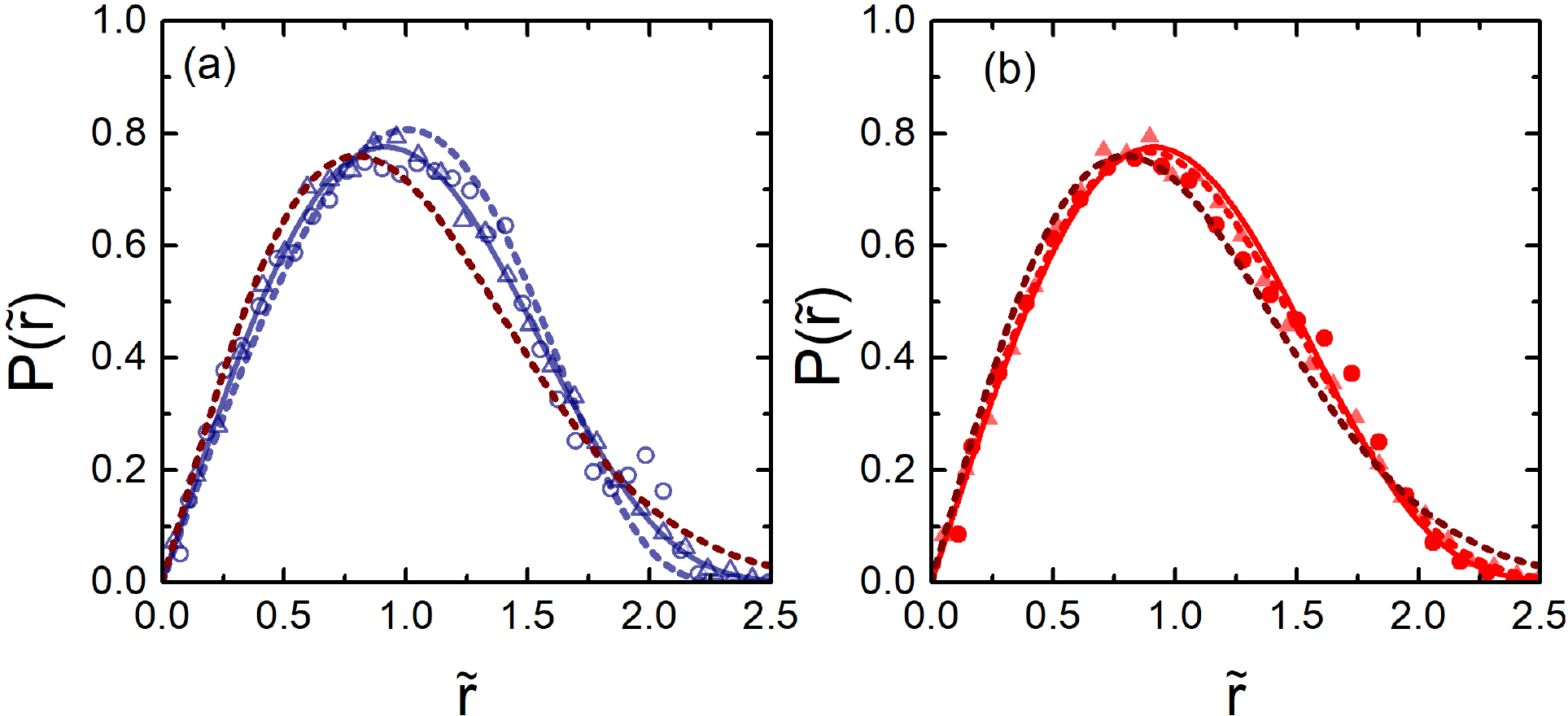}
	\caption{Distribution of the rescaled amplitudes $P(\tilde r)$ for the network with 1~dB (a) and 2~dB (b) attenuators (blue circles and red dots). They are compared to the analytical curves (blue and red dashed lines) and the RMT simulations (blue open and red full triangles) for $\gamma  =5.7$ and $\gamma  =12.8$, respectively. The maroon-dashed lines exhibit the bivariate Gaussian expected $P(\tilde r)$ in the Ericson regime.
}\label{Fig5}
\end{figure}

\begin{figure}[h!]
\includegraphics[width=0.6\linewidth]{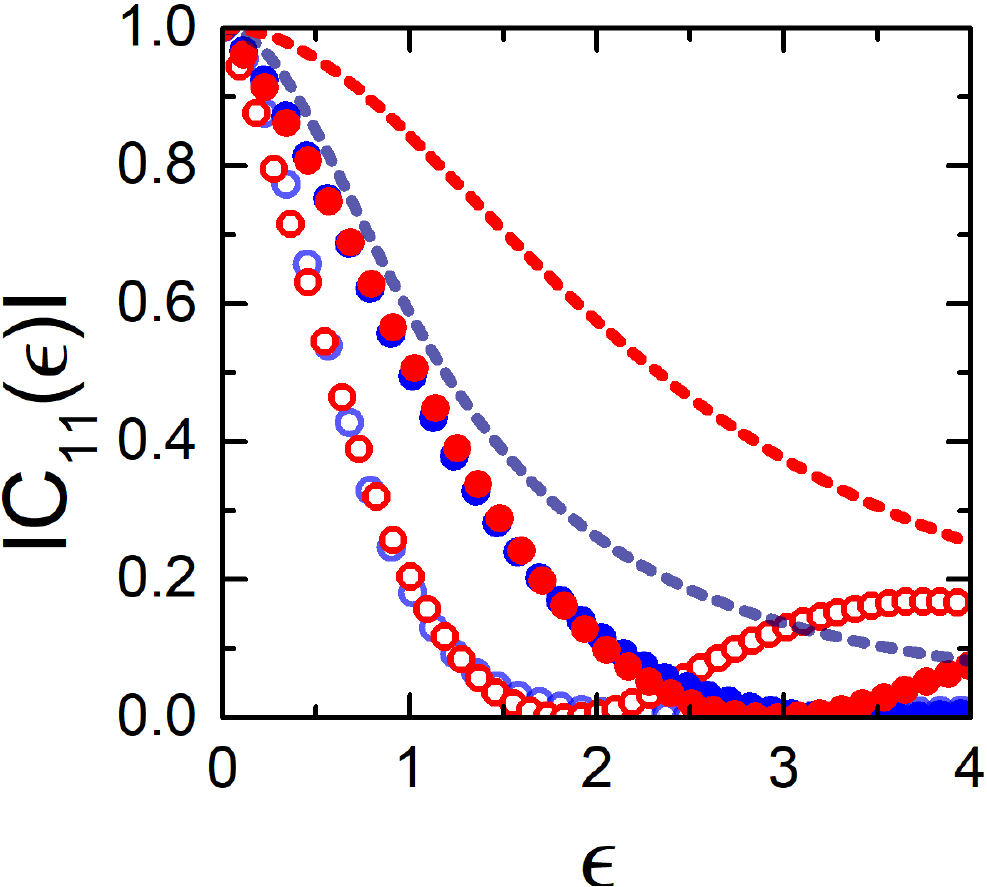}
\caption{
	Modulus of the two-point correlation function, $C_{11}(\epsilon)$, evaluated experimentally for the network with 1~dB (red circles) and 2~dB (red dots) attenuators is compared to the results obtained from the RMT simulations for $\gamma  =5.7$ (blue circles) and $\gamma  =12.8$ (blue dots), respectively. They are compared to the analytical curves for the same values of $\gamma$, shown as dashed lines of corresponding color.
}\label{Fig6}
\end{figure}

The distribution of the amplitudes $r=|S_{P\tilde iP\tilde i}|, \tilde i=1,1^\prime$ of the diagonal elements of the $S$ matrix is another important characteristic of quantum-chaotic scattering systems~\cite{Dietz2009,Dietz2010,Kumar2013,Kumar2017,Bialous2020,Chen2021}. It has not been investigated experimentally for quantum-chaotic systems belonging to the symplectic universality class. In~\reffig{Fig5} (a) and (b) we show the experimental distributions of the rescaled amplitudes, $P(\tilde r)$, (blue circles and red dots) for the networks with 1~dB and 2~dB attenuators, respectively. They are compared with the analytical results deduced from~\refeq{Eq.3} (solid lines) and the RMT simulations (blue open and red full triangles) for $\gamma  =5.7$ and $\gamma  =12.8$, respectively. In~\reffig{Fig6} we show the modulus of the two-point correlation function~\refeq{AutoCorr} of the $S$ matrix for the network with 1~dB (red circles) and 2~dB (red dots) absorption together with the RMT results for $\gamma  =5.7$ (blue circles) and $\gamma  =12.8$ (blue dots). Furthermore, we show the corresponding results for random matrices from the GUE as dashed lines of corresponding color. Agreement of the experimental results with the analytical curves and the RMT simulations for the GSE is good, whereas clear deviations from the GUE are observed for the distributions of the $K$ matrix, the reflection coefficients and the correlations functions. The rescaled distributions of the amplitudes also are well distinguishable for the case of 1~dB absorption, whereas for large absorption the GUE and GSE curves are close to each other. Note, that $\langle r\rangle $ and $\langle R\rangle$, and thus, $P(r)$ and $P(R)$ clearly differ for the GUE and GSE cases.
\section{Conclusions\label{Concl}} 
In summary, we performed experiments with open microwave networks with symplectic symmetry for intermediate and large loss parameters $\gamma = 5.7$ and $\gamma = 12.8$. Up to now, only the spectral properties of closed GSE networks have been investigated~\cite{Rehemanjiang2016,Rehemanjiang2018,Lu2020,Che2021} and shown to comply with GSE statistics except for deviations caused by nonuniversal contributions originating from short periodic orbits, and thus serve as a suitable model for experimental studies in the context of quantum chaos in systems with symplectic symmetry. 

As outlined in~\refsec{Sympl}, the $S$ matrix of a scattering system with symplectic symmetry is diagonal for two open channels. This manifests itself in the feature that in two-port measurements with microwave networks modelling quantum graphs with symplectic symmetry, transmission from one port to the other one is suppressed. Namely, the relative phase of $\pi$ in the two bonds that connect the subgraphs, which is required to achieve the symplectic symmetry, leads to a destructive interference of waves entering the network at one port and travelling through them, implying their suppression at the other port. Still, the microwave sent into the network at one port obviously visit both subgraphs, as can be concluded from the fact that their spectral properties agree well with those of random matrices from the GSE and clearly deviate from those of the GUE~\cite{Rehemanjiang2016,Rehemanjiang2018,Lu2020,Che2021}. Our aim was to find out, whether this still applies for the properties of the $S$ and $K$ matrix of an open GSE graph. We analyzed the distributions of the reflection coefficients and the real and imaginary parts of Wigner's reaction $K$ matrix with, respectively one port attached to corresponding vertices of the subgraphs. The results agree well with analytical results obtained within the framework of RMT for the $S$ matrices of corresponding dimension with symplectic symmetry and absorption and thus validate them. Similarly, we found good agreement between the experimental two-point $S$-matrix correlation function and the one obtained from Monte Carlo simulations based on the GSE. We come to the conclusion that the microwave networks, that are considered in this paper, indeed may serve, due to their simplicity also from the theoretical point of view, as an ideal test bed for open quantum-chaotic systems belonging to the symplectic universality class. Note, that the presence of short periodic orbits that are confined to a small fraction of the quantum graph -- a drawback for closed quantum graphs --, do not have any visible effects on the fluctuation properties of the scattering matrix associated with the corresponding open quantum graph.

\section{Acknowledgement}
This work was supported by the National Science Centre, Poland, Grant No. UMO-2018/30/Q/ST2/00324. BD thanks the National Natural Science Foundation of China under Grant No. 11775100, No. 11961131009, and No. 12047501 and acknowledges financial support from the Institute for Basic Science in Korea through the project IBS-R024-D1. 

\bibliography{References}
\end{document}